# Knowledge Distillation of Noisy Force Labels for Improved Coarse-Grained Force Fields

Feranmi V. Olowookere[1,4], Sakib Matin[2], Aleksandra Pachalieva[3], Nicholas Lubbers[1], and Emily Shinkle[1]

[1]*Computing and Artificial Intelligence Division, Los Alamos National Laboratory, Los Alamos, NM 87545, USA*

[2]*Theoretical Division, Los Alamos National Laboratory, Los Alamos, NM 87545, USA*

[3]*Earth and Environmental Sciences Division, Los Alamos National Laboratory, Los Alamos, NM 87545, USA*

[4]*The University of Alabama, Tuscaloosa, AL 35487, USA*

## Abstract

Molecular dynamics simulations are an integral tool for studying the atomistic behavior of materials under diverse conditions. However, they can be computationally demanding in wall-clock time, especially for large systems, which limits the time and length scales accessible. Coarse-grained (CG) models reduce computational expense by grouping atoms into simplified representations commonly called beads, but sacrifice atomic detail and introduce mapping noise, complicating the training of machine-learned surrogates. Moreover, because CG models inherently include entropic contributions, they cannot be fit directly to all-atom energies, leaving instantaneous, noisy forces as the only state-specific quantities available for training. Here, we apply a knowledge distillation framework by first training an initial CG neural network potential (the *teacher*) solely on AA-to-CG mapped forces to denoise those labels, then distill its force and energy predictions to train refined CG models (the *student*) in both single- and ensemble-training setups while exploring different force and energy target combinations. We validate this framework on a complex molecular fluid—a deep eutectic solvent—by evaluating two-, three-, and many-body properties and compare the CG and all-atom results. Our findings demonstrate that training a student model on ensemble teacher–predicted forces and per-bead energies improve the quality and stability of CG force fields.

## 1. Introduction

All-atom (AA) molecular dynamics (MD) simulations are widely used to probe structure, thermodynamics, and transport, [1-4] but generating long trajectories or large ensembles from these simulations can be computationally demanding in practice, especially for larger systems.[5] Coarse-grained (CG) models mitigate this cost by sacrificing fidelity, mapping groups of atoms into units (often called beads), thereby reducing both the number of particles and the number of interactions that must be computed.[6, 7] Additionally, the resulting variables typically evolve on a smoother potential energy surface (PES) than the underlying AA system, [8, 9] which in turn accelerates the exploration and sampling of the system's energy states.

CG force fields (FFs) are typically derived via either top-down or bottom-up protocols.[6, 9] The former approach chooses bead types and interaction forms to reproduce experimental observables or thermodynamic targets, [10, 11] but they may sacrifice structural fidelity; even widely used MARTINI-based models [11-13] can underpredict or misrepresent particle coordination when a single bead choice must best represent diverse local chemistries. On the other hand, the bottom-up approach derives effective interactions directly from AA data by matching forces or distribution functions. [14-17] Yet, they are state-dependent and struggle to capture many-body effects arising from the averaged-out AA degrees of freedom. [9, 17, 18] More recently, data-driven CG models have demonstrated that learning flexible, many-body bead interactions can improve accuracy and transferability compared to pairwise CG approaches. [7, 19-22] Although, faithfully reproducing dynamical properties remains an open challenge; we refer the reader to recent efforts in improving CG dynamics.[23-27]

Machine learning (ML) methods for CG have lately advanced alongside ML-based development of AA force fields. [28-30] In particular, architectures originally designed for AA potentials [31] naturally extend from atom-wise contributions to the system energy into bead-wise contributions to the CG free energy. It is important to note that a bottom-up CG model is bound to the underlying accuracy of its AA FF, so any ML workflow should yield consistent performance regardless of which AA reference is used.

Bottom-up CG training includes two important and related challenges. First, projecting AA forces $\boldsymbol{f}(\boldsymbol{r})$ from AA positions $\boldsymbol{r}$ onto bead forces $\boldsymbol{F}(\boldsymbol{R})$ over bead positions $\boldsymbol{R}$ requires averaging over many AA microstates (consistent with the same CG configuration), thereby smoothing out detailed fluctuations. However, the AA-to-CG mapped force data obtained from AA MD retains unsmoothed, noisy labels.[9, 32] This noise does not arise from stochastic simulation effects, since AA MD is deterministic, but rather from the intrinsic conditional variance associated with averaging over multiple AA microstates for a given CG configuration. Hence, one must sample enough CG configurations (> $10^6$ snapshots total) to ensure that the error is dominated by variations in the CG PES rather than by noise in the force labels. [20] Second, the effective CG energy functions are potentials of mean force (PMF), a form of free energy, and therefore include entropic contributions that are not tractable to compute. Although one can include AA energies in a CG loss function, doing so yields poor CG results, so in practice only the instantaneous, noisy force labels are used for training the model. [17]

To address these limitations, knowledge distillation (KD) [33] provides a training paradigm in which an initial *teacher* model guides a more accurate *student* model by supplying auxiliary targets.[33-35] In other words, combining noisy labels with the teacher's denoised outputs enhances the accuracy of the student. Originally applied to image classification, [33] KD has recently been adapted in molecular simulations to enhance the accuracy and efficiency of ML interatomic potentials.[36-39] Distillation from an ensemble of teachers further reduces variance and yields smoother, more reliable targets than any single model, [40-42] which is valuable especially when labels are noisy. To our knowledge, KD has not been applied to the CG domain, where supervision is dominated by noisy projected forces and where intermediate energy signals could be particularly beneficial; building CG models is highly related to the task of de-noising, [43] and models of intractable free energies can be built using derivative information alone. [44] Related ML approaches [45-48] reduce CG–AA differences by using contrastive or likelihood-free objectives that implicitly denoise the training signal, or by matching selected observables or probability distributions instead of the projected forces.

Here, we introduce a KD-based workflow for ML CG potentials using the Hierarchically Interacting Particle Neural Network with Tensor Sensitivity (HIP-NN-TS) architecture, [49, 50] recently shown to construct thermodynamically transferable CG models.[19] The crucial observation behind our work is that initial teacher models perform estimation of the denoised forces and forms an implicit integration for the intractable free energy, and that these quantities can be used as auxiliary targets for second-generation models to improve bottom-up CG training. We validate our framework on a complex molecular fluid, namely, a deep eutectic solvent (DES), by first training the teacher only on AA-to-CG mapped forces extracted from AA simulations. We then train student models on those AA-to-CG mapped forces along with various combinations of teacher-generated targets—forces, per-bead energies, and system energy. We show that students distilled from an ensemble of teachers and supervised with teacher forces along with per-bead energies achieve better accuracy of structural properties while maintaining single-model inference speed.

## 2. Method

Our simulation workflow for training and validating teacher and student models is outlined in Figure 1. We begin by generating AA data with MD simulations and mapping this data to a CG representation. Next, we train the teacher networks on this mapped AA data and subsequently train student networks on the same reference set augmented with auxiliary targets (forces and energies) generated by the teachers.

**Generation of AA data for CG training:** We generated AA data by simulating a system of 1,000 DES molecules composed of 250 choline, 250 chloride, and 500 urea molecules using the GROMACS package. [51] Bonded and non-bonded interactions were modeled with the Optimized Potentials for Liquid Simulations–All-Atom (OPLS-AA) force field parameterized for DESs. [52, 53] The initial configuration was built with Packmol, [54] using a cubic box of 44.84 Å that corresponds to the experimentally determined density of 1.196 g/cm$^3$. [55] This was followed by energy minimization using the steepest-descent algorithm. We then equilibrated for 5 ns in the canonical (NVT) ensemble at 298.1 K with a Nosé-Hoover thermostat (0.5 ps time constant). [56] Following equilibration, we performed a 1 ns production run in the NVT ensemble and saved snapshots every 1 ps, resulting in 1,000 configurations. Throughout equilibration and production, we used a 2 fs time step with periodic boundary conditions applied in all directions; hydrogen bond lengths were constrained with Linear Constraint Solver (LINCS) algorithm. [57] Dispersive and electrostatic interactions were truncated at 10 Å and long-range electrostatics were computed via the Particle Mesh Ewald method.[58]

The AA trajectories were mapped to the CG representation by applying our bead-mapping protocol (discussed below) to AA positions and forces, producing the dataset used to train ML CG models.

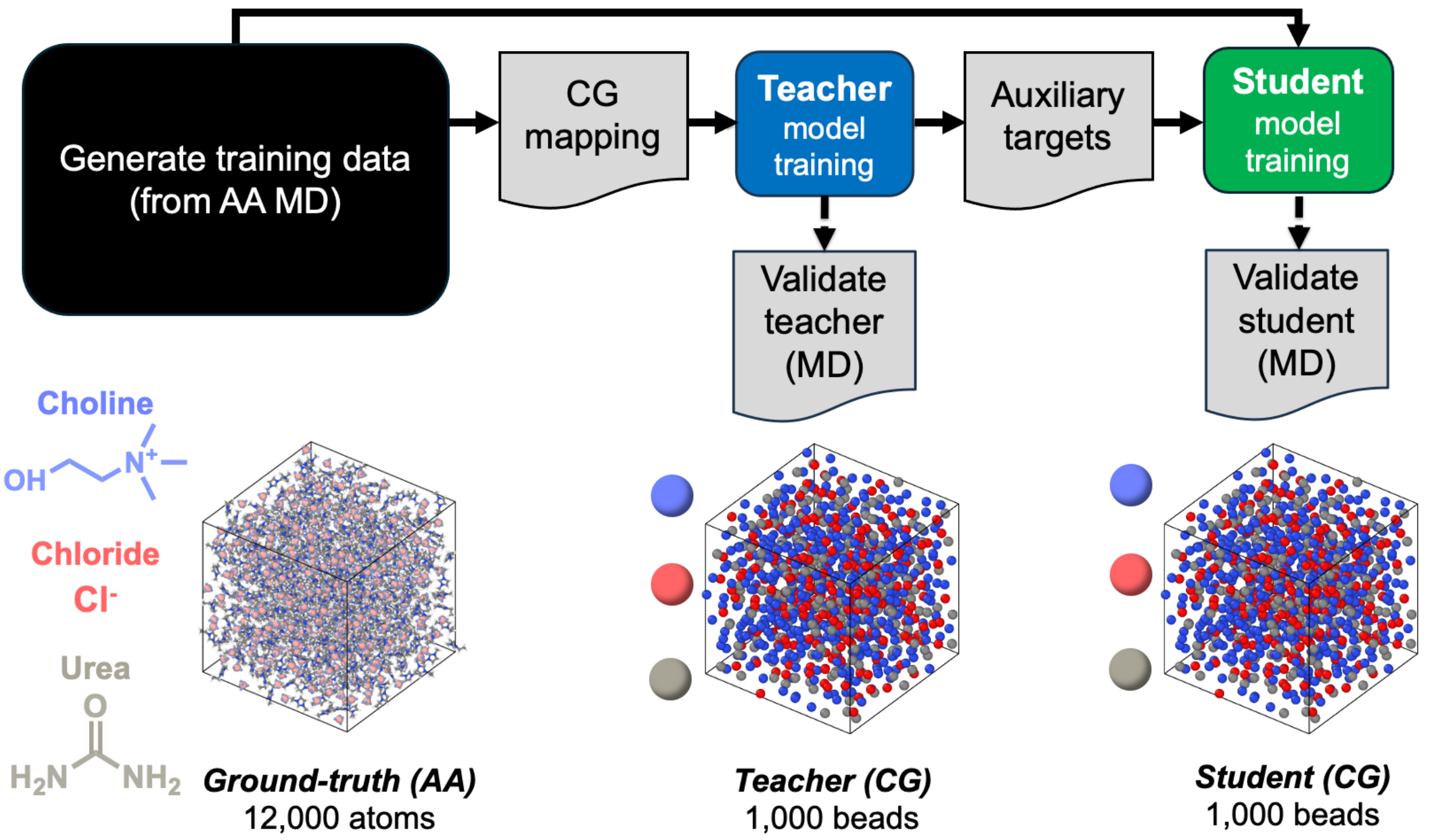

**Figure 1.** Simulation workflow for training and validating teacher and student models. MD: molecular dynamics, AA: all-atom, CG: coarse-grained. Each molecule is represented by one bead at the coarse-grained level.

**Coarse-Graining Scheme:** Following our previous work, [19] we use a bottom-up strategy to the CG force field.

$$\langle -\nabla U_{AA} \rangle = -\nabla U_{CG} \quad (1)$$

We define the CG free-energy surface, or PMF ($U_{CG}$) in Eq. 1, so that it matches the AA PES ($U_{AA}$) over a representative range of configurations in phase space, and we require the CG forces derived from this surface to reproduce the AA force averages conditioned on each CG configuration. In this study, each molecule is represented by a single bead; the DES therefore contains three bead types: choline (Cho), chloride (Cl), and urea, as shown in Figure 1.

CG positions are computed as centers of mass of the atoms assigned to each bead, and each bead mass is the sum of the atomic masses in the corresponding molecule. The force on a CG bead is taken as the sum of the atomic forces within that bead, which provides an unbiased estimator of the negative gradient of the CG free energy with respect to the bead coordinates. Although alternative mapping schemes exist, [48] the force-mapping operator must remain consistent with the chosen coordinate mapping; specifically, after contraction over atomistic indices, it should act as the inverse of the AA Jacobian of the coordinate map. Throughout this study, we treat the AA-to-CG mapped data obtained from the AA trajectory as the “ground-truth” for training.

**Model architecture:** For both teacher and student CG models, we use the HIP-NN-TS, [49, 50] a graph convolutional neural network that represents the system energy as a sum of per-bead contributions computed through a hierarchy of interaction and bead-environment layers. It extends the original HIP-NN model [49] by incorporating a tensor sensitivity component, enabling each neuron in the interaction layer to encode many-body features. The convolutions operate on pairwise displacement vectors and thus naturally produce invariance to rotations, translations, and permutations of identical beads. The network featurizes each bead's local environment using its type and neighbor displacement vectors, then predicts energy contributions at each hierarchy which are summed to yield per-bead energies $\varepsilon_i$ and the system energy $E$. Forces on each bead are then obtained by automatic differentiation of $E$ with respect to the bead positions.

**Table I**. HIP-NN-TS hyperparameters.

| Parameter | Symbol | Value |
|---|---|---|
| Tensor order | $\ell$ | 2 |
| Interaction layers | $n_{int}$ | 1 |
| Atomic-environment layers | $n_{atom}$ | 3 |
| Sensitivity functions | $n_v$ | 20 |
| Atomic features per layer | $n_{feature}$ | 32 |

The HIP-NN-TS hyperparameters in the ML CG models are shown in Table I, selected by trial and error in our previous work [19] and have performed well in prior HIP-NN-TS applications to atomistic simulations. [36, 49] To ensure stability in sparsely sampled regions, we supplement the learned potential with a physics-based short-range repulsive term. This additional pairwise potential is activated whenever two beads approach closer than a cutoff $r_0$, defined by the radial distribution function (RDF), thereby preventing unphysical overlaps and avoiding extrapolation of the ML component into untrained, small-$r$ regimes.

$$E_{rep}(r) = E_0 e^{-ar} \tag{2}$$

The repulsive term takes the form in Eq. 2, where $r$ is the inter-bead distance and $E_0$, $a > 0$ are system-specific parameters.[19] Example training scripts for both teacher and student models are available in the open-source hippynn repository.[49]

**Teacher–Student training overview:** We train one or more teacher models on the noisy ground-truth forces $\boldsymbol{F}_i$ and then use the trained model(s) to generate auxiliary targets. These signals are appended to the initial training set to guide the learning of an improved (more accurate and/or faster) student model. The teacher-provided supervision captures richer structure in the data; in this study, it includes per-bead energies $\varepsilon_i$, total molecular energy E, and forces $\boldsymbol{f}_i$. We train the student in two settings: using the auxiliary targets (or knowledge) from a single teacher or from an ensemble of eight teachers. Although ensemble size can be varied, we used eight for all experiments; preliminary tests with four teachers produced similar results, thus eight was chosen to ensure robust averaging throughout this study.

**Teacher model training:** Eight teacher ($T$) models were trained on the same dataset using an identical architecture and size, differing only by their random initialization seeds. The teachers are trained on the AA-to-CG mapped dataset $\mathcal{D} : \{(\boldsymbol{R}_i, Z_i)\} \longmapsto \{\boldsymbol{F}_i\}$ where $\boldsymbol{R}_i$ are $i$th bead positions (computed as centers of mass of the atoms assigned to each bead), $Z_i$ are bead types, and $\boldsymbol{F}_i$ are bead forces. Training uses stochastic gradient descent to minimize the loss function:

$$\mathcal{L}_{teacher} = w_F \mathcal{L}_{err}(\widehat{\boldsymbol{F}_i}, \boldsymbol{F}_i) \tag{3}$$

The error term in Eq. 3 is defined as an equal-weight sum of root mean squared error (RMSE) and mean absolute error (MAE), which has proved successful for training previous HIP-NN-TS models [19, 36]

$$\mathcal{L}_{err}(\hat{y}, y) = RMSE(\hat{y}, y) + MAE(\hat{y}, y) \tag{4}$$

Where $\hat{y}$ denotes the model prediction and $y$ the target in Eq. 4. In addition to using the eight independently trained teachers ($T$) during MD inference, we also deploy their averaged ensemble, denoted as *T8*. After training, we extract $\varepsilon$, $E$, and $\boldsymbol{f_i}$, to construct an augmented dataset for student model training.

**Variants of knowledge distillation:** We train the student in two regimes: *S1* (supervision from a single teacher *T*) and *S8* (ensemble-averaged supervision from eight teachers *T8*). Student variants are defined by their auxiliary targets in Table II, varying the force target (ground-truth $\boldsymbol{F_i}$ only, teacher $\boldsymbol{f_i}$ only, or both) and the energy target (per-bead energy $\varepsilon_i$ only, system energy $E$ only, or both). Student model names follow the convention of the training regime (*S1* or *S8*), followed by the included targets. For example, *S1Fε* denotes a student trained with single-teacher guidance using losses on ground-truth forces and teacher-provided per-bead energies only. As another example, *S8FfEε* denotes a student trained with ensemble guidance using losses on all the targets.

Table II lists only a subset of all target combinations because we restricted the experiment matrix to the questions of greatest interest. Beginning with $\varepsilon$ as the baseline, we toggled the force inputs $\boldsymbol{F_i}$ and $\boldsymbol{f_i}$, to identify the best force configuration and teacher type; we then fixed those choices and varied the energy targets ($E$ and/or $\varepsilon_i$), to probe the most informative cases while keeping the number of training runs manageable.

**Table II.** Variants of Knowledge Distillation for Student Models.

| Variant | Teacher type | System energy $E$ | Per-bead energy $\varepsilon$ | Denoised forces $\boldsymbol{f_i}$ | AA forces $\boldsymbol{F_i}$ |
|---|---|---|---|---|---|
| *S1Fε* | Single | ✗ | ✓ | ✗ | ✓ |
| *S1fε* | Single | ✗ | ✓ | ✓ | ✗ |
| *S1Ffε* | Single | ✗ | ✓ | ✓ | ✓ |
| *S8Ff* | Ensemble | ✗ | ✗ | ✓ | ✓ |
| *S8Fε* | Ensemble | ✗ | ✓ | ✗ | ✓ |
| *S8fε* | Ensemble | ✗ | ✓ | ✓ | ✗ |
| *S8Ffε* | Ensemble | ✗ | ✓ | ✓ | ✓ |
| *S8FfE* | Ensemble | ✓ | ✗ | ✓ | ✓ |
| *S8FfEε* | Ensemble | ✓ | ✓ | ✓ | ✓ |

**Student model training:** Depending on the target, each student variant is trained on the augmented dataset $\mathcal{D} : \{(\boldsymbol{R_i}, Z_i)\} \longmapsto \{\boldsymbol{F_i}, \boldsymbol{f_i}, E, \varepsilon_i\}$ (where $i$ represents bead index) with the loss function:

$$\mathcal{L}_{student} = \mathcal{L}_{teacher} + w_f \mathcal{L}_{err}\left(\boldsymbol{f}_i^S, \boldsymbol{f}_i^T\right) + w_E \mathcal{L}_{err}(E^S, E^T) + w_\varepsilon \mathcal{L}_{err}\left(\varepsilon_i^S, \varepsilon_i^T\right) \quad (5)$$

The alignment terms $\mathcal{L}_{err}\left(\boldsymbol{f}_i^S, \boldsymbol{f}_i^T\right)$, $\mathcal{L}_{err}(E^S, E^T)$, and $\mathcal{L}_{err}\left(\varepsilon_i^S, \varepsilon_i^T\right)$ in Eq. 5 encourage the student's forces $\boldsymbol{f}_i^S$, per-bead energy partition $\varepsilon_i^S$, and system energy $E^S$ to match those of the teacher's $\boldsymbol{f}_i^T$, $\varepsilon_i^T$, and $E^T$, respectively. It is important to note that we normalized the system-energy loss by the number of beads to prevent it from dominating the loss function relative to the other loss terms. To ensure a controlled comparison, both teacher and student models have the same architecture and size. All other settings and hyperparameters in Table I remained fixed, and the students differed only by the addition of the alignment term during training.

**Table III.** Loss weights used for different target types in student models.

| Target type used | System energy weight $w_E$ | Per-bead energy weight $w_{\varepsilon_i}$ | Denoised forces weight $w_{f_i}$ | AA forces weight $w_{F_i}$ |
|---|---|---|---|---|
| *S1Fε* | - | 5 | - | 1 |
| *S1fε* | - | 5 | 1 | - |
| *S1Ffε* | - | 5 | 1 | 2 |
| *S8Ff* | - | - | 1 | 2 |
| *S8Fε* | - | 5 | - | 1 |
| *S8fε* | - | 5 | 1 | - |
| *S8Ffε* | - | 5 | 1 | 2 |
| *S8FfE* | 5 | - | 1 | 2 |
| *S8FfEε* | 5 | 5 | 1 | 2 |

In Table III, when only per-bead energies were used, the energy-loss weights were set to $w_{\varepsilon_i} = 5$ and $w_E = 0$. When only the system energy was used, they were set to $w_{\varepsilon_i} = 0$ and $w_E = 5$. When both per-bead and system energies were included, both energy-loss weights were set to $w_{\varepsilon_i} = w_E = 5$. In variants that include both $\boldsymbol{F_i}$ and $\boldsymbol{f_i}$, the force-loss term was partitioned as $w_{F_i} = 2$, $w_{f_i} = 1$, thereby emphasizing true forces while still leveraging the teacher's denoising signal. Preliminary tests on these weights suggest that the chosen values yield stable optimization and faster convergence during training. We also found that scaling by factors of 2–5 produced similar validation trends, whereas larger imbalances caused the energy terms to overwhelm force learning despite normalization.

**Model Validation and Analyses:** We use MD to validate model performance via the Large-Scale Atomic/Molecular Massively Parallel Simulator (LAMMPS), [59] averaging results from eight independent replicas per model, with error bars denoting one standard deviation. Each replica was initialized with a configuration drawn from the training dataset and unique velocity seed, equilibrated for 1 ns, then sampled every 5 ps to yield 100 frames for model validation via the TRAVIS package. [60]

During MD with the eight-teacher ensemble (*T8*), we evaluate all eight networks at each time step and update the positions with their averaged force. On the other hand, the distilled student (*S8*) is trained on these ensemble-averaged forces and energies, so it needs only a single network evaluation per step during inference. Equilibration was assessed by inspecting the potential-energy time series (see Figure S1) with the pymbar steady-state heuristic. [61] Both *T* and *S1* models reached equilibrium by 400 ps, whereas the *T8* and *S8* variants equilibrated very rapidly (< 2 ps).

We validated each CG model by comparing its simulation output with statistics from the reference AA trajectories. Because the CG training data contain noisy forces and lack explicit energy labels, regression metrics such as MAE, RMSE, and coefficient of determination ($R^2$) are unreliable; consequently, we judge model quality primarily through structural distribution functions. Specifically, we computed three complementary metrics to probe two-body, three- body, and many-body interactions. Dynamics were not examined here because CG models inherently sample more rapidly than AA counterparts due to the smoothed PES. Although methods [24-27] such as including frictional forces [25] or applying scaling relationships [23, 62] have been proposed, addressing dynamics falls outside this study's scope.

Two-body structure was quantified using the RDF,

$$g(r) = \frac{\langle N(r) \rangle}{4\pi r^2 \Delta r \rho} \tag{6}$$

which measures the probability of finding a particle at distance $r$ from a reference particle, where $\Delta r$ is the bin width, $\langle N(r) \rangle$ is the ensemble average number of particles found between $r + \Delta r$, $\rho = \frac{N}{V}$ is the bulk number density with $N$ particles in volume $V$, and $4\pi r^2 \Delta r$ is the spherical shell volume. To follow the standard notation $g(r)$ in Eq. 6, it is worth noting that we use $r$ here as the inter-bead distances for both AA and CG RDFs.

Three-body structure was quantified by the angle distribution function (ADF). For each triplet $(i, j, k)$ with central atom $j$, we compute

$$\theta_{ijk} = \cos^{-1}(\widehat{\boldsymbol{R}}_{ji}, \widehat{\boldsymbol{R}}_{jk}) \tag{7}$$

and accumulate

$$P(\theta) = \frac{1}{N_{triplet}} \langle \sum_{j=1}^{N} \sum_{i \neq j} \sum_{\substack{k \neq j \\ k \neq i}} \delta(\theta - \theta_{ijk}) \rangle \tag{8}$$

subject to the neighbor-cutoff conditions $|\widehat{\boldsymbol{R}}_{ji}|$, $|\widehat{\boldsymbol{R}}_{jk}| < R_{max}$ (6.0 Å, 6.5 Å, or 7.5 Å). These values were chosen because they span the location of the first-solvation-shell peak in the corresponding RDFs, ensuring that the ADF captures the primary local structure. In Eq. 8, $N_{triplet}$ is the total number of $(i, j, k)$ triplets satisfying those cutoffs, and $\langle \cdots \rangle$ denotes an ensemble average.

To quantify many-body aggregation, we follow the hierarchical clustering procedure of Frömbgen et al, [63] where two beads $i$ and $j$ belong to the same cluster when their separation $R_{ij} = \|R_i - R_j\|$ is below a chosen cutoff $R_{cut}$ ($R_{cut} \leq \frac{L_{box}}{2}$ for a cubic box). Starting at $R_{cut} = 0$ (all beads are isolated), we increase the cutoff continuously. Each time the criterion $R_{ij} = R_{cut}$ is met for the first time, two existing clusters merge; the corresponding cutoff value is stored as a merge distance $d_m$. Plotting the histogram of all merge distances,

$$C(R_{cut}) = \sum_m \delta(R_{cut} - d_m) \tag{9}$$

gives the cluster distribution function (CDF) in Eq. 9: a profile of how frequently clusters form as $R_{cut}$ increases, reflecting the structural heterogeneity of the system.

To quantify deviations between the CG distributions $t_1(R)$ and the ground-truth AA distributions $t_2(R)$ (whether RDFs, ADFs, or CDFs), we employ a finite-sum approximation of the total absolute error (TAE): [19]

$$TAE(t_1, t_2) \approx \sum_{i=0}^{n} |t_2(R_i) - t_1(R_i)| \Delta R \tag{10}$$

In Eq. 10, $\{R_i\}_{i=0}^{n}$ is a uniform grid of distances from $R_0 = 0$ to $R_n = R_{max}$, $\Delta R = R_{i+1} - R_i$ is the bin width, $t_1$ and $t_2$ denote the two probability densities (either $g(r)$, $P(\theta)$, or $C(R_{cut})$).

## 3. Results and Discussion

In this section, we report training metrics for our ML CG models and highlight the challenge of training teacher models solely on the initial dataset that contains only noisy forces. We then examine how single-teacher supervision compares with an ensemble of teachers when training student models, and we evaluate performance in both training and inference. Finally, we study the effect of different force and energy targets to identify an effective protocol for training the student.

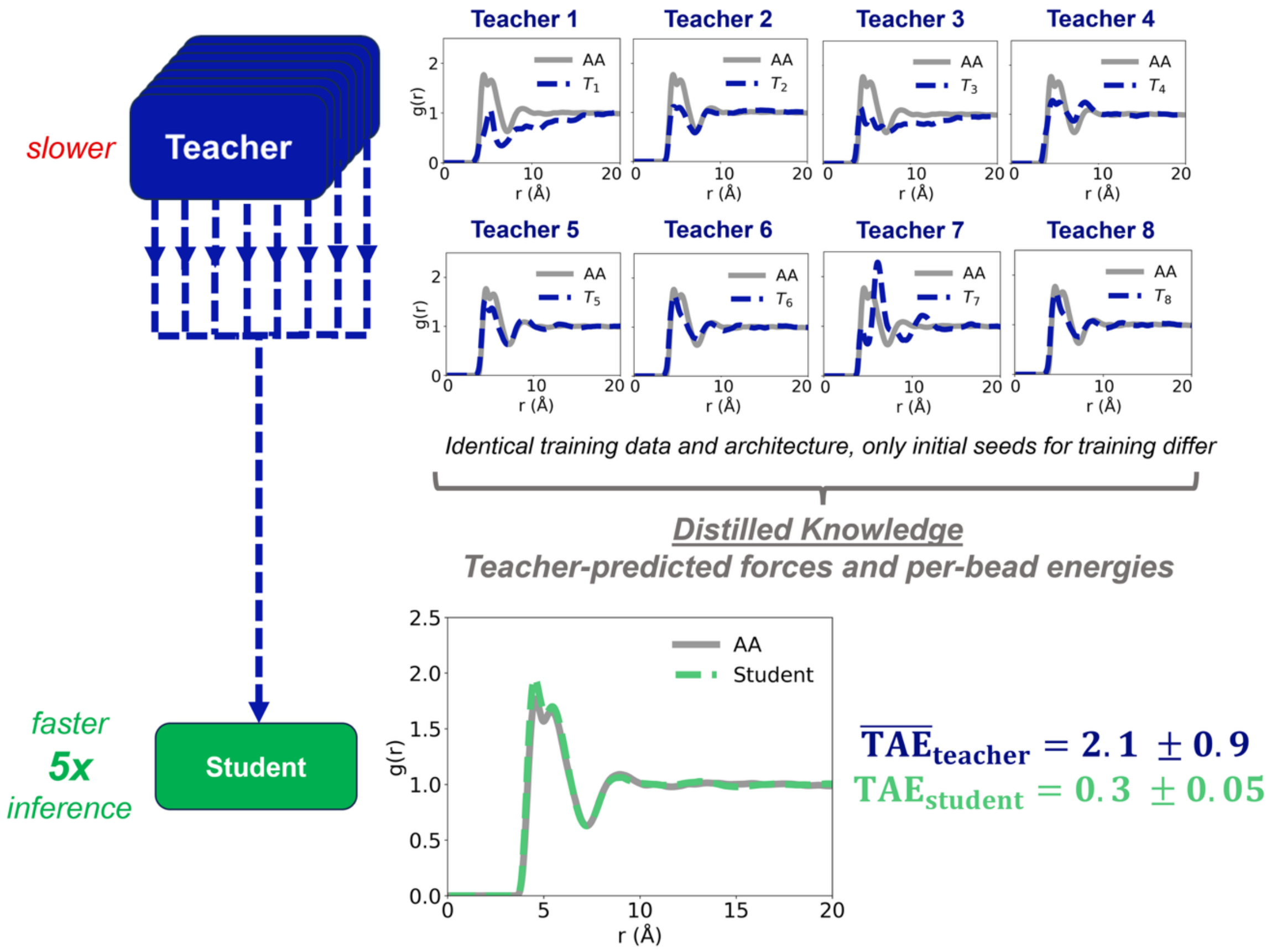

**Figure 2.** Distilling auxiliary targets (or knowledge) from an ensemble of teacher models into a single student improves both accuracy and efficiency of ML CG models. Each teacher is trained on the same AA force data but with different random seeds; averaging their predictions yields denoised forces and per-bead energies, which are then combined with the original training data to train the student. While the teachers exhibit bias in the RDF compared to the AA reference, the student both reproduces the reference RDF accurately and achieves roughly fivefold faster inference than the teacher-ensemble (*T8*) model. Note: RDFs shown here are that of Choline-chlorine pairs in the system.

Figure 2 previews our main findings. Despite identical architectures, single-teacher models with different initial training seeds produced highly variable, unstable dynamics with clustering. However, distilling the average predictions (forces and per-bead energies) of an ensemble of teachers into a single student improves the stability and accuracy, while delivering $\sim 5\times$ faster inference than the teacher ensemble. We quantify these trends below in training and inference metrics.

### 3.1 Training Metrics

A common approach to evaluating ML models is to compare their predictions against the true values from held-out test data during training. For our teacher models, Figure 3 shows that across different random seeds, the MAE, RMSE, and coefficient of variation $R^2$ metrics exhibit little variation. Additionally, each teacher attains only $R^2 \approx 0.35$, which is low compared to typical ML benchmarks ($R^2 > 0.9$). This outcome is expected, since the instantaneous forces $\boldsymbol{F}_i$ are drawn from a distribution which can be quite wide, [48] whereas the network learns to estimate the conditional mean $\langle \boldsymbol{F}_i \rangle$, inherently adding noise to the loss.

Figure S3 presents the corresponding metrics for the student variant S8, distilled from the ensemble of eight teachers. Here, the force-prediction $R^2$ changes only slightly ($0.35 < R^2 < 0.38$) with the choice of energy targets. Likewise, the energy-prediction $R^2$ (for $\varepsilon_i$ and $E$) remains nearly the same ($R^2 > 0.98$) except in cases where that quantity is omitted from training, in which case its $R^2$ drops.

Overall, while MAE, RMSE, and $R^2$ provide useful quantitative benchmarks, they alone are insufficient to assess ML CG model quality. Accordingly, we place greater emphasis on distributional and structural comparisons in the analyses that follow.

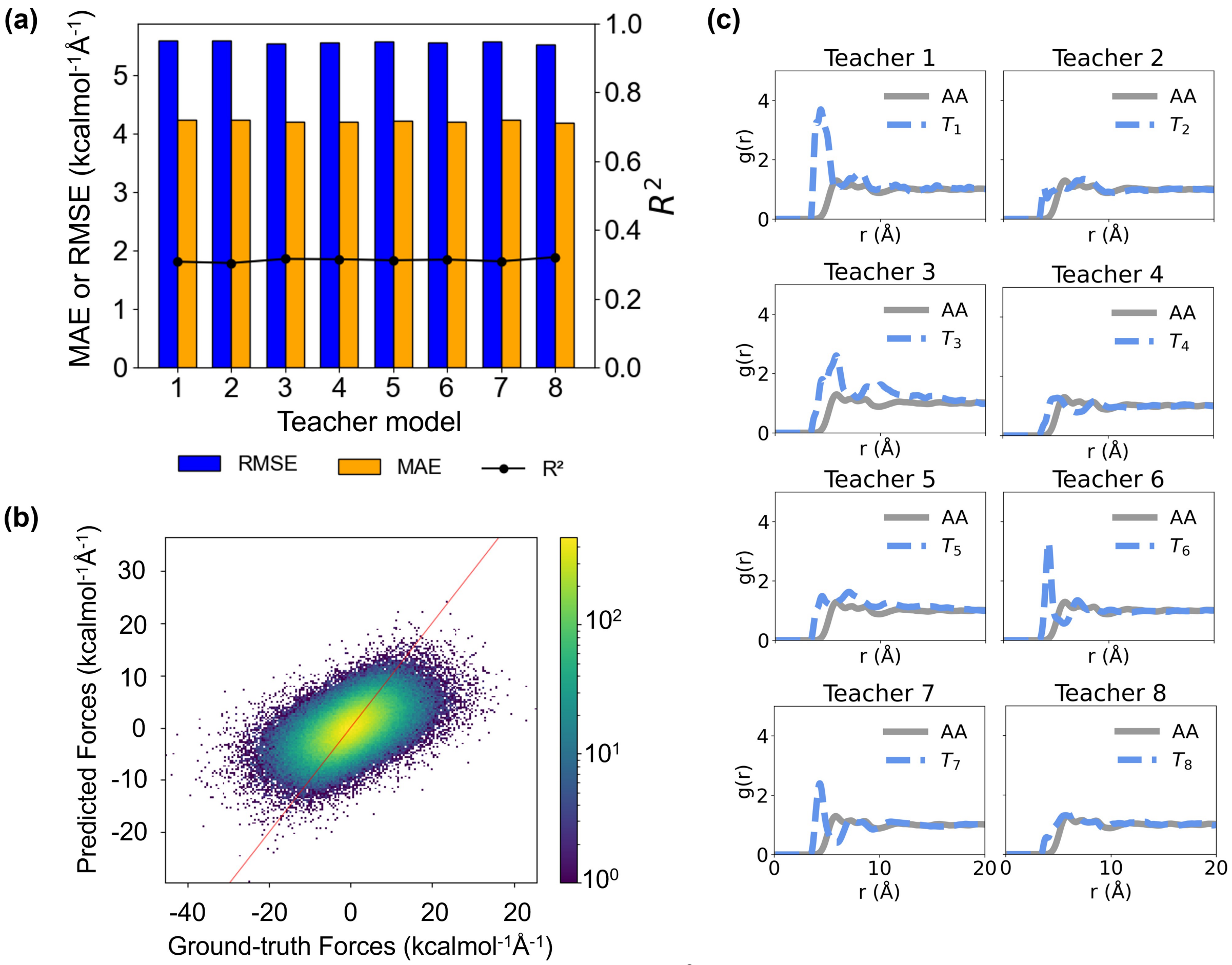

**Figure 3.** (a) Training metrics MAE, RMSE and $R^2$ of individual teacher models on ground-truth AA force targets, (b) Parity plot of the predicted versus ground-truth AA forces, (c) Urea-Urea RDF of individual teacher models in comparison to AA reference.

## 3.2 Effects of teacher source in student training

We trained eight teacher models (denoted as $T$), all using the same architecture and dataset but initialized with different random seeds. They produce noticeably different outcomes: for example, in Figure 3, Teacher 1 exhibits a shorter-range urea–urea peak near 5 Å with markedly greater intensity (4× higher), whereas Teacher 8 more closely matches the AA distribution.

The individual RDFs and corresponding RDF TAEs for each $T$ model are provided in Section S3.1 of the Supplementary Material, while the individual CDFs and CDF TAEs are shown in Section S5.1. Figure 4 summarizes the RDF TAE for the urea–urea pair, with summary RDF TAEs for the remaining pairs reported in Section S3.3 of the Supplementary Material. Similarly, summary ADF and CDF TAEs are shown in Figures 5 and 6, respectively, with the full ADF and CDF results available in Sections S4.1 and S5.3 of the Supplementary Material.

In Figure 4, the average RDF TAE of the $T$ model is large across all pairs and feature wide error bars, indicating high variability among teachers; as a rule of thumb, RDF TAE > 1 signifies substantial distortion of pair structure. In Figure 5, an ADF TAE greater than 0.2° indicates a significant deviation from the reference; by this measure, the $T$ models exhibit high ADF TAEs for most triplets except Cl–Cl–Cl. ADF TAEs decrease with increasing angular cutoff $r_{max}$, because larger cutoffs sample more triplets and average out random fluctuations. In Figure 6, CDF TAEs are similarly high, particularly for urea, with CDF TAE > 0.5 indicating pronounced CDF deviations.

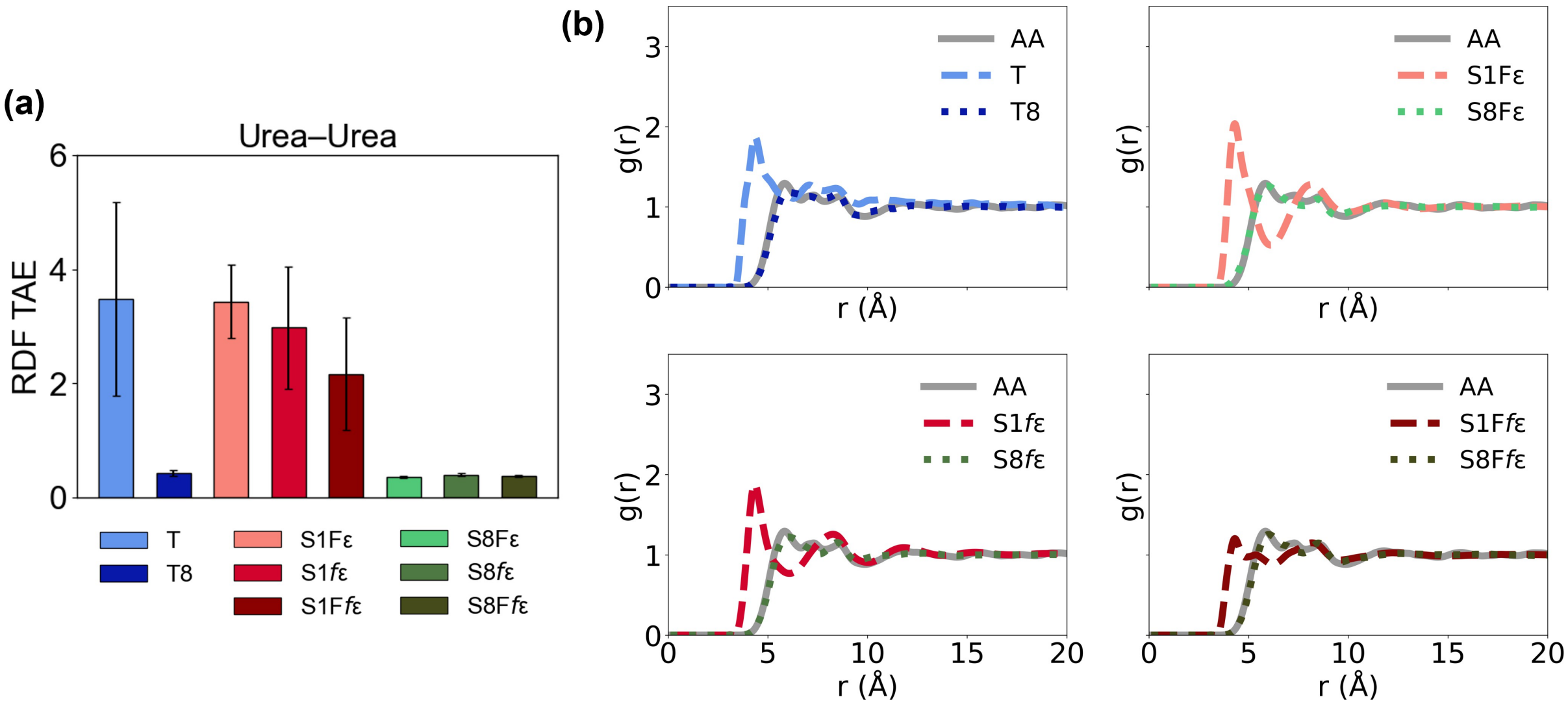

**Figure 4.** Comparison of urea-urea (a) RDF TAE and (b) RDF for teacher and student models (using different force targets) relative to the AA reference. Error bars denote one standard deviation over 8 replicas. Regarding energy data inclusion, all student models here were trained to only per-bead energies. Model naming follows the convention in Table II: $\boldsymbol{F}$, ground-truth forces; $\boldsymbol{f}$, teacher forces; $E$, system energy; $\varepsilon$, per-bead energies.

To identify the origin of these errors, we calculated time-resolved CDFs for each species: choline CDFs are shown in Figure 7, while those of chloride and urea in Figure S2. The chloride CDF remains essentially stable over 0–300 ps. In contrast, the choline and urea CDFs change significantly, with their dominant peaks shifting to smaller cutoff distances, which reflects progressive clustering during the simulation. These results indicate that teachers trained only on ground-truth forces may lack sufficient accuracy and stability to represent the CG system. This likely explains the prolonged potential-energy equilibration time (~ 500 ps) even when starting from an equilibrated AA-to-CG mapped configuration.

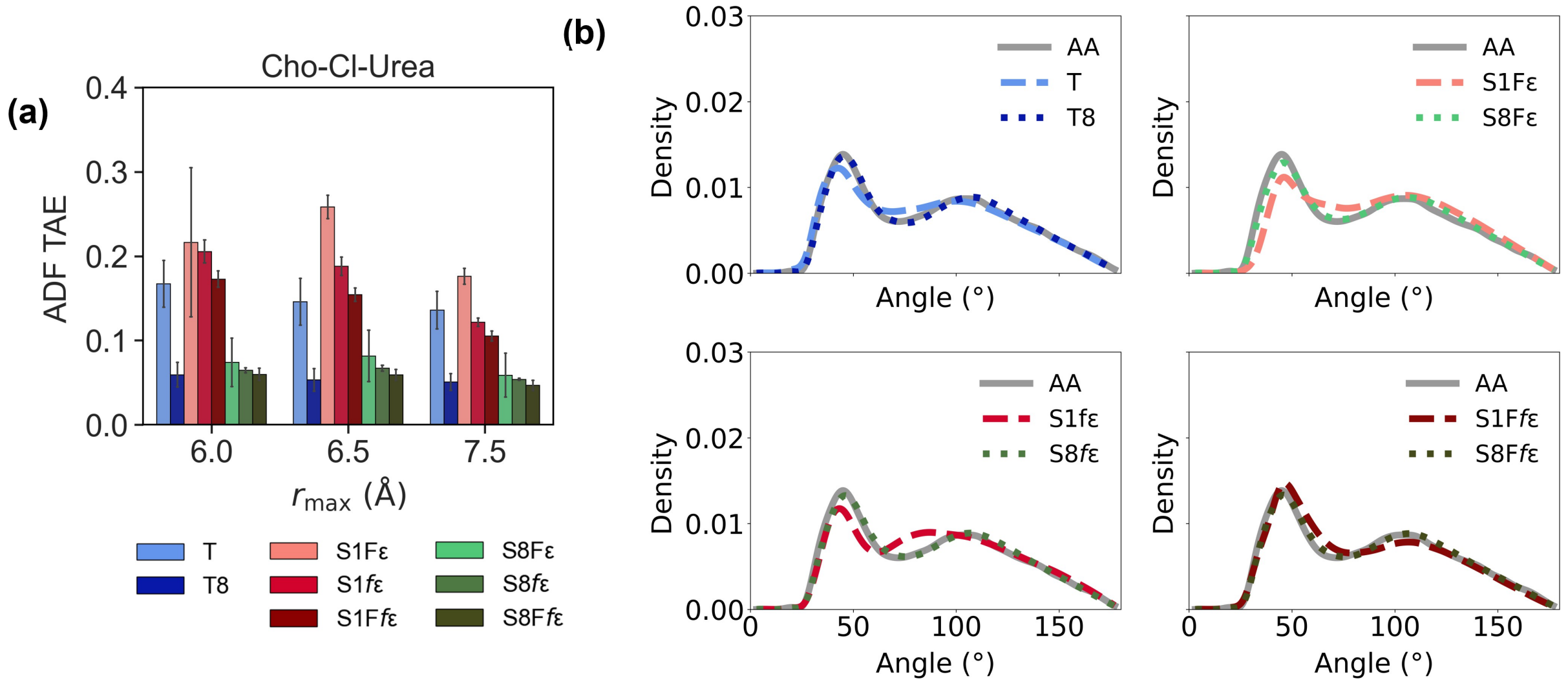

**Figure 5.** Comparison of Cho-Cl-Urea (a) ADF TAE at different ADF cutoff values $r_{max}$ and (b) example ADF at $r_{max}$ = 7.5 Å at for teacher and student models relative to the AA reference (using different force targets). Error bars denote one standard deviation over 8 replicas. Regarding energy data inclusion, all student models here were trained to only per-bead energies. Model naming follows the convention in Table II: $\boldsymbol{F}$, ground-truth forces; $\boldsymbol{f}$, teacher forces; $E$, system energy; $\varepsilon$, per-bead energies.

When we instead directly deploy the eight-teacher ensemble (denoted as *T8*) in the inference stage, the RDF, ADF, and CDF TAEs all drop substantially compared to single-teacher runs, and the simulations remain stable (no spurious clustering or time-dependent drift in the CDFs). Although all teacher models are trained on the same data, each may learn a slightly different approximation to the CG mean force. [64] Averaging over these realizations in *T8* could mitigate this realization dependence to yield a more robust approximation to the CG mean force.

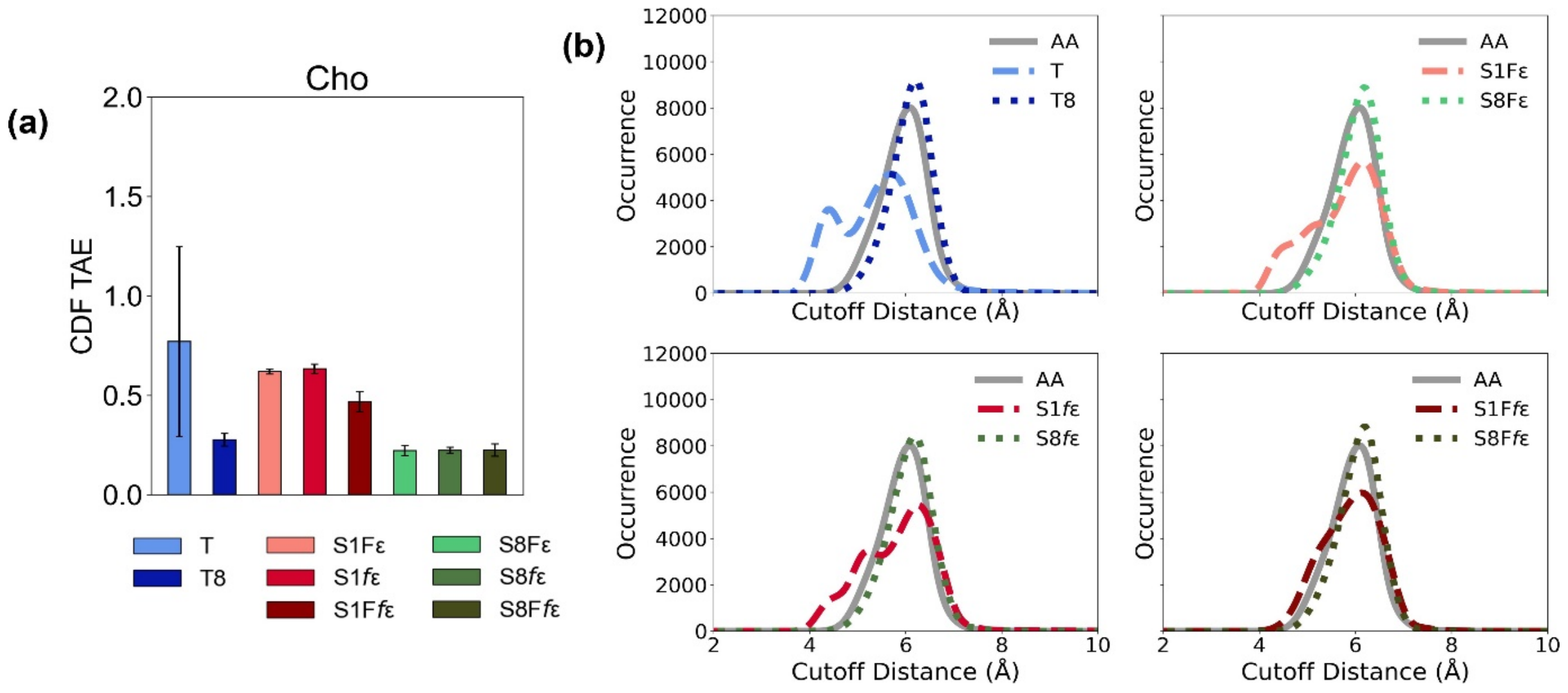

**Figure 6.** Comparison of Choline (Cho) (a) CDF TAE and (b) CDF for teacher and student models (using different force targets) relative to the AA reference. Error bars denote one standard deviation over 8 replicas. Regarding energy data inclusion, all student models here were trained to only per-bead energies. Model naming follows the convention in Table II: $\boldsymbol{F}$, ground-truth forces; $\boldsymbol{f}$, teacher forces; $E$, system energy; $\varepsilon$, per-bead energies.

We next trained two sets of student models: *S1*, supervised by the *T* model, and *S8*, supervised by the *T8* model. We computed the same structural metrics and compared them to the AA reference. In Figures 4–6, *S1* results are plotted in red shades and *S8* in green shades. The *S1* models closely mirror the single-teacher behavior—their RDF, ADF, and CDF TAEs match or exceed those of *T* (for example, the urea CDF TAE in Figure 6 is higher than the corresponding *T* value)—and visual inspection confirms similar clustering and aggregation. This behavior likely reflects that each *T* model may provide a biased approximation to the CG mean force, and *S1* learns these biases, leading to similar or slightly amplified TAEs.

By contrast, *S8* achieves TAEs comparable to *T8* and yields stable dynamics, with time-invariant CDFs for all species (Figure 7). Across Figures 4, 5, S20, and S39, the *T8* model generally performs on par with the *S8* variants and in some cases shows marginally lower TAEs: for example, for the Cho–Cl–Urea ADF in Figure 5 and the Cl–Cl–Cl ADF in Figure S20. However, these differences are consistently small (typically $< 0.05°$), as reflected by the overlapping error bars.

**Table IV.** Performance of model types on a 1000-bead DES system at 300 K. All benchmarks used a 2 fs time step on a 48-thread CPU.

| Model type | Performance (ns/day) |
|---|---|
| Single e.g., *T, S8* | 2.66 |
| Ensemble e.g., *T8* | 0.54 |

The primary advantage of the distilled *S8* model is that it maintains the accuracy gained by using an ensembled model, without incurring the additional computational burden that ensemble evaluation requires. Essentially, it shares the same architecture and size as *T*/*T8,* but requires only one model evaluation per MD step, whereas *T8* must evaluate eight teachers and average their force outputs during MD integration. This makes *S8* run approximately five times faster than *T8* in our benchmarks.

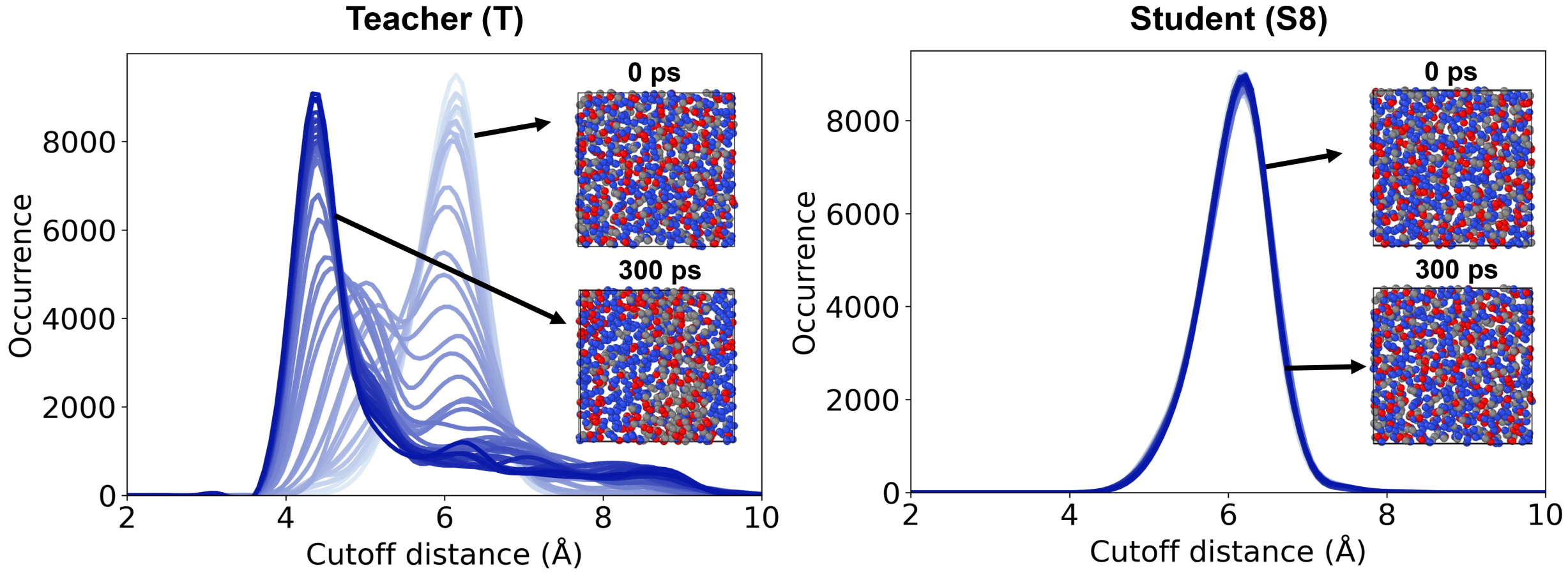

**Figure 7.** Temporal evolution of Choline CDFs for the teacher and ensemble-trained student models, shown at 10 ps intervals from 0 ps (lightest blue) to 300 ps (darkest blue). The teacher's distribution gradually shifts, reflecting species aggregation, while the student models' curves remain nearly unchanged, indicating stability over time.

Moreover, the additional cost of training *S8* is minimal. Deploying the *T8* ensemble already requires training eight teacher models; the only extra step introduced by distillation is training a single student model using the ensemble-generated auxiliary targets. Overall, training the student on ensemble guidance improves both the accuracy and efficiency of ML CG simulations.

For comparison with other simulation approaches, ML AA MD provides a more appropriate baseline than classical CG FFs (e.g., Martini), since our models employ more complex architectures and therefore naturally incur higher per-step costs. [65] In our previous study, [19] we reported 30–150× speedups relative to ML AA MD after accounting for both reduced cost in evaluating forces and larger integration time steps. Actual performance, however, also depends on design choices, including network architecture, model size, training hyperparameters, and the mapping scheme.

### 3.3 Effects of force targets

To determine an optimal student-training protocol, we first varied the force supervision: ground-truth AA forces only (*S1Fε*, *S8Fε*), teacher-predicted forces only (*S1fε*, *S8fε*), and both (*S1Ffε*, *S8Ffε*); it is important to note that only $\varepsilon_i$ was included as an energy target in all cases here. Performance was evaluated via RDF, ADF, and CDF TAEs in Figures 4–6 (full summary TAEs are in the Supplementary Material), with individual RDFs in Section S3, ADFs at $r_{max}$ = 6.0, 6.5, 7.5 Å in Section S4, and species CDFs in Section S5 of the Supplementary Material.

For the *S1* variants, using only teacher-predicted forces generally reduces the RDF TAE compared to using only AA forces. The lowest RDF TAEs are achieved when both force targets are combined, though the error bars indicate this improvement is modest. CDF TAEs follow the same pattern. ADF TAEs also improve with teacher forces except for Cl–Cl–Cl and Cho–Cho–Cho at $r_{max}$ = 6.0 Å, likely due to limited sampling.

For the *S8* variants, adding teacher forces (*S8Ffε*) has minimal effect on RDF and CDF TAEs relative to *S8Fε*, while ADF TAEs decrease slightly across most triples, except that of Cho–Cho–Cho which remained largely unchanged.

In summary, including teacher-predicted forces could benefit student accuracy for both single- and ensemble-teacher training.

### 3.4 Effects of energy targets

To isolate the effects of energy-based auxiliary targets, we evaluated four *S8* variants: without energy targets (*S8Ff*), with per-bead energies only (*S8Ffε*), with system energy only (*S8FfE*), and with both per-bead and total energies (*S8FfEε*). Regarding force targets, all models here were trained using both $\boldsymbol{F_i}$ and $\boldsymbol{f_i}$ for consistency. We only focus on *S8* models here because they have been shown earlier to outperform *S1* models.

Figure 8 reports the RDF TAE for the Cho–Cl pair; complete RDF, ADF, and CDF profiles and their corresponding TAEs for all energy-target models are provided in Sections S6–S8 of the Supplementary Material. Unlike the student models shown in Figures 4–6 (which all already included $\varepsilon_i$), Figure 8 explicitly varies the inclusion of energy targets to assess model performance. Thus, *S8Ffε* is the only model common to both Figure 8 and Figures 4–6.

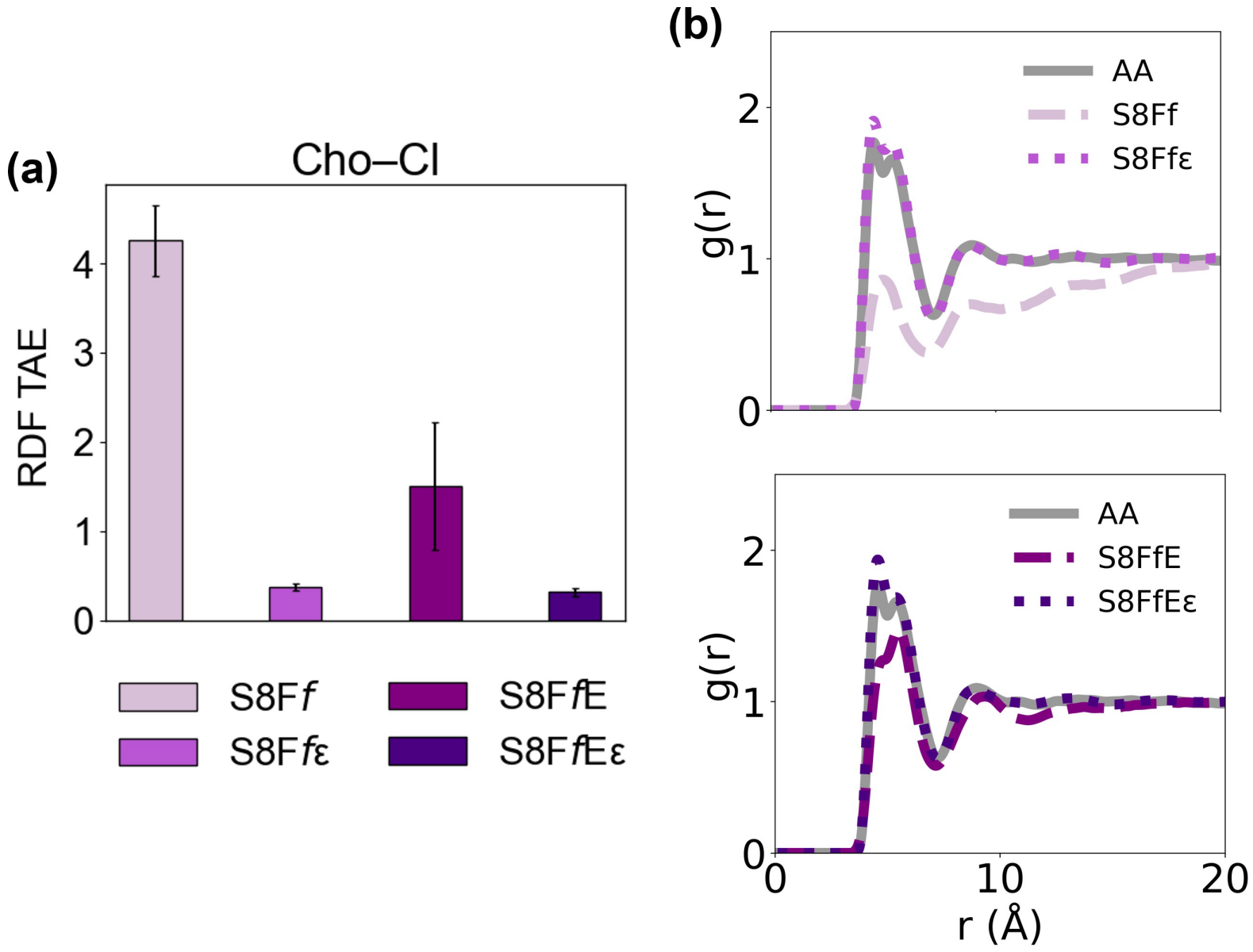

**Figure 8.** Comparison of Choline(Cho)-Chloride(Cl) (a) RDF TAE and (b) RDF for student models using different energy targets relative to the AA reference. Error bars denote one standard deviation over 8 replicas. *S8* models were trained on averaged data from all eight teachers. Regarding force data inclusion, all models here were trained to both ground-truth and teacher forces. Model naming follows the convention in Table II: $\boldsymbol{F}$, ground-truth forces; $\boldsymbol{f}$, teacher forces; $E$, system energy; $\varepsilon$, per-bead energies.

Among the models shown in Figure 8, the absence of energy targets i.e., *S8Ff* leads to the largest RDF TAE, whereas including per-bead energies i.e., *S8Ffε* substantially reduces the error and yields accurate, stable distributions across all species. Using only the total system energy results in slightly higher TAEs than *S8Ffε*, while combining both per-bead and system energies restores performance to the level of *S8Ffε*. The same qualitative trends are observed for ADF and CDF TAEs (Supplementary Material Sections S6–S8). These comparisons imply that the strong performance of the *S8* students in Figures 4–6 depends critically on $\varepsilon_i$: without it, the *S8* accuracy is significantly reduced, and *T8* would indeed be the most accurate model.

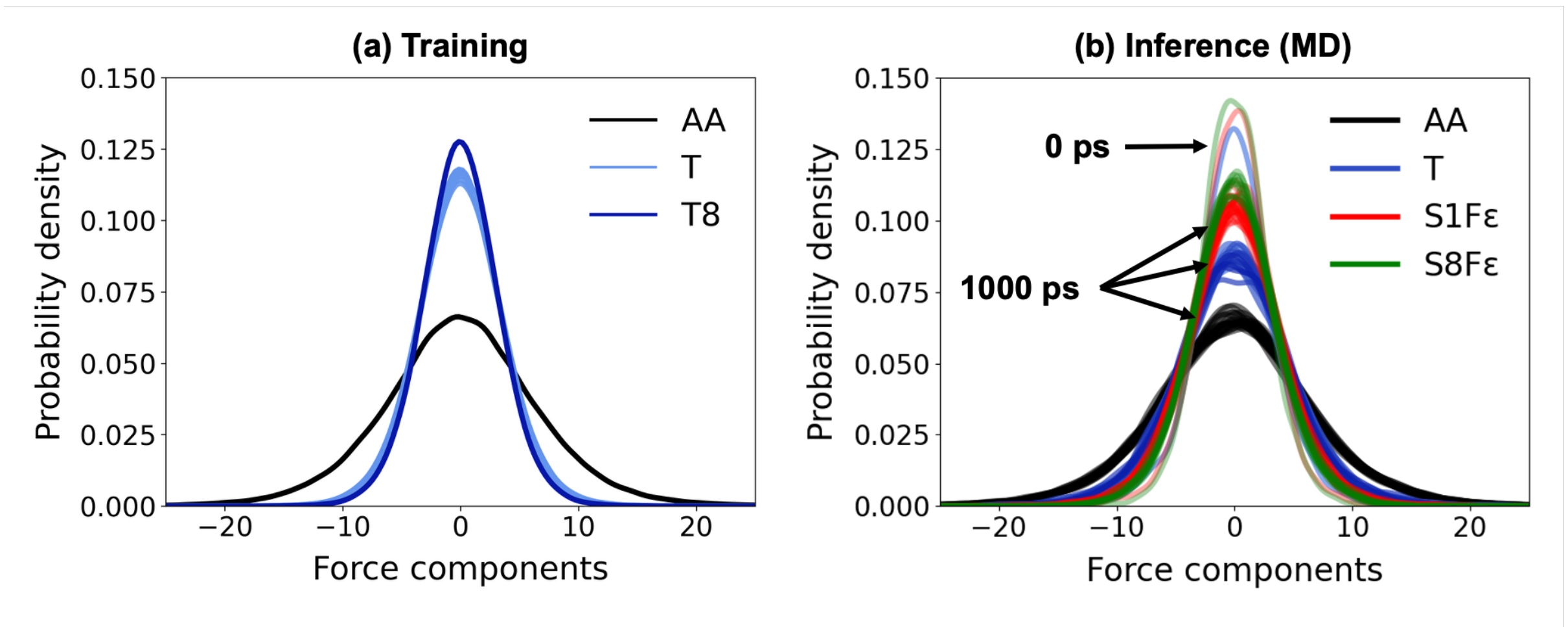

**Figure 9.** (a) Cartesian force distributions for AA training data and for forces predicted by the *T* and *T8* teacher models, which are used to train the *S1* and *S8* students, respectively. (b) Cartesian force distributions sampled during MD with the AA, *T*, *S1Fε* and *S8Fε* models.

Shading in (b) indicates simulation time from 0 ps (lightest) to 1000 ps (darkest). In both panels, AA represents the AA-to-CG mapped forces.

Before training the *S1* and *S8* models, *T* and *T8* in Figure 9a predict nearly identical force distributions and both are much narrower than the ground-truth (AA) force distribution, which shows broader tails due to projection noise (see Figure 3). However, during MD, the force distribution in Figure 9b generated by *T* broadens and develops heavier tails, becoming closer to the AA distribution, whereas *S1Fε* gets narrower and *S8Fε* exhibits the narrowest distribution. All ML models begin with similar force distributions, as shown by the lighter curves, but diverge as MD progresses; after this initial transient, the distributions remain stable over the 1000 ps trajectory. These results suggest that knowledge distillation improves the stability of force statistics under self-consistent sampling.

In Figure 8, $\varepsilon_i$ supervision allows a student model to recover much of the accuracy typically achieved by an ensemble, without incurring the associated computational overhead at inference. It is thus evident to conclude that $\varepsilon_i$ is the key target for stability and accuracy; incorporating $E$ in addition offers little benefit, though it does not degrade performance when used alongside $\varepsilon_i$.

In summary, our results suggest that ensemble-guided training with auxiliary energy supervision, optionally using teacher-predicted forces, provides effective regularization relative to noisier AA-to-CG mapped force labels. The narrower force distributions reflect a more robust approximation to the conditional mean force, analogous to noise-minimizing force mapping strategies, [48] and may contribute to their improved stability under on-the-fly sampling.

## 4. Conclusion

Here, we present a knowledge distillation workflow for ML CG FFs to overcome noisy force labels and the difficulty of directly fitting CG energies to AA references due to its inherent inclusion of entropic contributions. Specifically, we trained initial teacher CG neural network potentials solely on those instantaneous, noisy forces. We then distilled both force and energy predictions from the teachers into student CG models and evaluated all models by comparing RDFs, ADFs, and CDFs against the AA reference using the TAE metric, as standard training metrics (MAE, RMSE, $R^2 \approx 0.35$) proved unreliable.

We trained eight teacher models with identical architecture and dataset, varied only by training seed. We observed that all teachers (in MD simulations) caused species aggregation that distorted structure, which produced large RDF, ADF, and CDF TAEs. Instead, deploying their ensemble in MD produced stable simulations with improved structural accuracy. We then trained student models on auxiliary targets from either a single teacher or an eight-teacher ensemble. Students distilled from one teacher retained its poor performance, while the ensemble-distilled student matched the ensemble's stability and accuracy at single-model speed, roughly five times faster.

We also tested various student training protocols and found that incorporating teacher-predicted forces yields a small accuracy gain. The most impactful auxiliary target is the per-bead energy, while including system energy has little effect. Training students on ground-truth AA forces together with per-bead energies and forces predicted by the teacher ensemble therefore improves the quality of CG FFs.

Future work will apply this framework to more challenging materials such as polymers, whose high configurational variability produces much noisier forces. We will also explore successive student generations, training each new model on the auxiliary targets of its predecessor, although recent studies [40] suggest little gain beyond the first generation. We also plan to evaluate how network size and architecture affect the trade-off between accuracy and speed. While our main aim here is to improve CG model stability and accuracy under noisy forces, future work may also address network compression. We anticipate that this approach can deliver accurate, efficient CG FFs to study phenomena at higher length and time scales.

**Supplementary Material Section**

Temporal evolution of model observables, training metrics, two- (RDF), three-(ADF) and many-(CDF) body evaluations of teacher and student models with varying force and energy targets, input scripts for reproducing workflow.

**Acknowledgments**

Research presented in this article was supported by the National Security Education Center (NSEC) Informational Science and Technology Institute (ISTI) using the Laboratory Directed Research and Development program of Los Alamos National Laboratory project number 20240479CR-IST as part of the Applied Machine Learning Summer Research Fellowship program. This research used resources provided by the Darwin testbed at LANL which is funded by the Computational Systems and Software Environments subprogram of LANL's Advanced Simulation and Computing program. LANL is operated by the Triad National Security, LLC, for the National Nuclear Security Administration of the U.S. Department of Energy (contract no. 89233218NCA000001). We are grateful to Heath Turner and Galen Craven for insightful discussions and suggestions. We acknowledge the use of AI writing tools to aid with phrasing and proofreading of the manuscript.

**Notes**

The authors declare that they have no known competing financial interests or personal relationships that could have appeared to influence the work reported in this paper.

**Author Information**

**Corresponding Authors**

- Emily Shinkle - Computing and Artificial Intelligence Division, Los Alamos National Laboratory, Los Alamos, New Mexico 87545, United States; Email: eshinkle@lanl.gov
- Nicholas Lubbers - Computing and Artificial Intelligence Division, Los Alamos National Laboratory, Los Alamos, New Mexico 87545, United States; Email: nlubbers@lanl.gov

**Authors**

- Feranmi V. Olowookere – Department of Chemical and Biological Engineering, The University of Alabama, Tuscaloosa, AL 35487-0203, United States.
- Sakib Matin - Theoretical Division, Los Alamos National Laboratory, Los Alamos, New Mexico 87545, United States.
- Aleksandra Pachalieva - Earth and Environmental Sciences Division, Los Alamos National Laboratory, Los Alamos, New Mexico 87545, United States; Center for Nonlinear Studies, Los Alamos National Laboratory, Los Alamos, New Mexico 87545, United States.

**CRediT Contributions**

- F.V.O. Software, validation, investigation, visualization, data curation, writing—original draft.
- S.M. Conceptualization, methodology, validation, writing—review and editing.
- A.P. Conceptualization, validation, writing—review and editing.

- N.L. Conceptualization, validation, methodology, software, supervision, project administration, resources, Funding acquisition, writing—review and editing.
- E.S. Conceptualization, validation, methodology, software, supervision, project administration, resources, Funding acquisition, writing—review and editing.

**Data and Software Availability**

We provide all simulation files used in this study: GROMACS inputs for training-data generation and LAMMPS scripts for CG simulations are listed in Section S9 of the SI. The FF and structure files are available in the OPLS-DES repository, [52] and the model-training scripts can be found in the open-source hippynn repository (https://github.com/lanl/hippynn).

**Highlight Image**

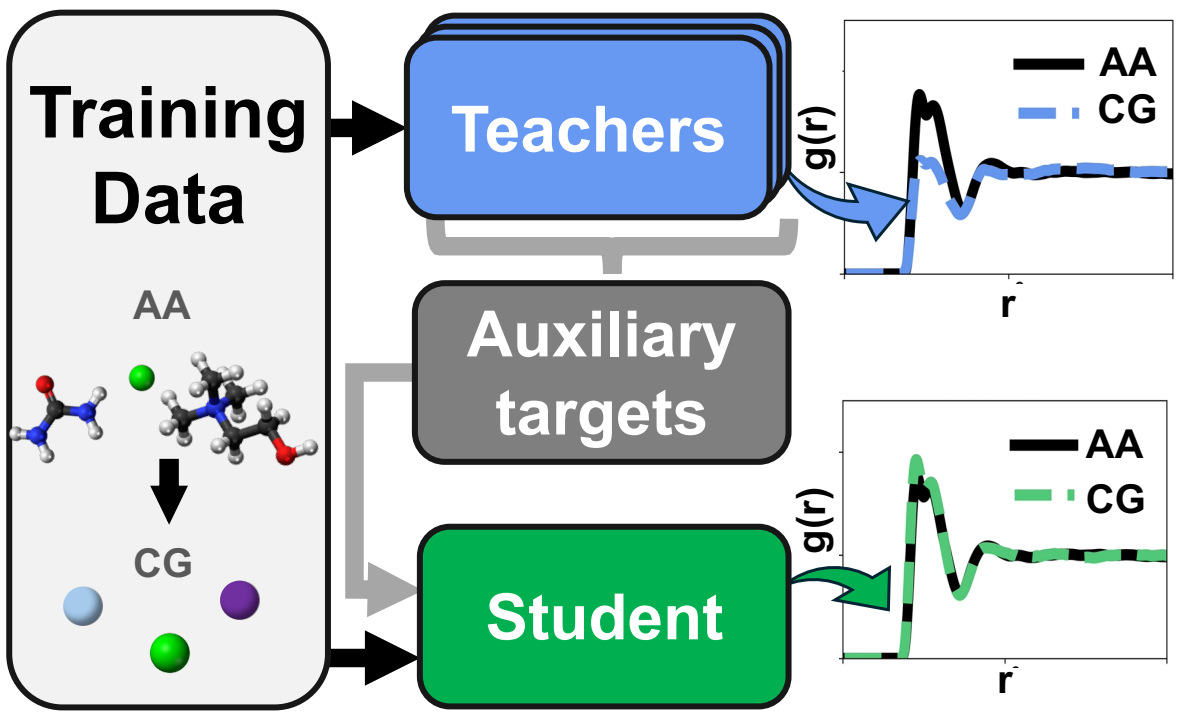